\documentclass[reprint,aps,prd,showpacs,footinbib,amsmath,amssymb,amsfonts,superscriptaddress]{revtex4-1}
\usepackage{marginnote}
\usepackage{hyperref}
\usepackage[section]{placeins}
\usepackage{amssymb}
\usepackage{bm}
\usepackage{verbatim}
\usepackage{latexsym}

\usepackage{lipsum}
\usepackage{mathtools}
\usepackage{amsfonts,amssymb}
\usepackage{graphicx,epsfig}
\usepackage{psfrag}
\usepackage{amsmath,amssymb}
\usepackage{mathrsfs}

\usepackage{bm}
\usepackage{latexsym}
\usepackage{mathrsfs}

\newcommand{\addlabel}[1]{%
\refstepcounter{equation}%
\addlabel{#1}%
\let\]\endequation }

\def\MPlh{M_{_{\rm Pl}}}

\begin{document}

%%%
\title{Bounce inflation driven by Higgs field}

%%%%
\author{Jose Mathew}\email{josecherukara@gmail.com}
\affiliation{School of Physics, Indian Institute of Science Education and Research Thiruvananthapuram (IISER-TVM), Trivandrum 695551, India}
%\affiliation{School of Physics, Indian Institute of Science Education and Research Thiruvananthapuram (IISER-TVM), Trivandrum 695016, India}

%\affiliation{ School of Physics, Indian Institute of Science Education and Research Thiruvananthapuram (IISER-TVM), Trivandrum 695016, India}
%\author{Jose Mathew}\email{jose@iisertvm.ac.in}
%\affiliation{ School of Physics, Indian Institute of Science Education and Research Thiruvananthapuram (IISER-TVM), Trivandrum 695016, India}
%\author{Joseph P Johnson} \email{joseph@iisertvm.ac.in}
%\author{S. Shankaranarayanan} \email{shanki@iisertvm.ac.in}

%\affiliation{School of Physics, Indian Institute of Science Education and Research Thiruvananthapuram (IISER-TVM), Trivandrum 695016, India}

\begin{abstract}
In this work, we investigate a model of bounce inflation driven by the Higgs field. The Higgs field is non-minimally coupled with gravity through the Gauss-Bonnet term. We show that the Higgs field could drive a power-law contraction followed by a bounce and thereafter an inflationary phase with exit. The phases of contraction and inflation are obtained analytically. The smooth transition to the inflationary phase from contraction is obtained numerically. Further, the power-spectrum of the model is found to be consistent with the cosmological data.
\end{abstract}

\maketitle
%\vskip{2cm}

\section{Introduction}
Inflationary paradigm~\cite{book:15483,book:15485,book:15486,book:75690,mukhanov1992theory} introduced in the early 1980's solves the puzzles associated with the hot Big Bang Cosmology. Most importantly, it also provides a successful mechanism for the generation of primordial perturbations. Its ability to explain the origin of structure and anisotropies in Cosmic Microwave Background~(CMB) makes it the best paradigm describing early stages of the Universe. From current PLANCK data~\cite{ade2014planck}, we know that the temperature fluctuations in CMB are nearly scale-invariant. Hence inflationary models can be scrutinized and constrained for its ability to provide an explanation for this near scale-invariance~\cite{lyth1999particle,lyth2008particle,mazumdar2010particle,yamaguchi2011supergravity} . However, there is no unique mechanism for inflation, and there are plenty of models that satisfy the observations by providing the observed spectrum. This together with the fact that inflation requires new physics makes the problem unsettled. Recently there is a growing interest in alternatives to inflation such as bounce, bounce inflation, etc~\cite{brandenberger2009alternatives}.

Cosmological bounce~\cite{novello2008bouncing,cai2014exploring,lilley2015bouncing,ijjas2016implications,brandenberger2017bouncing} is a paradigm in which the universe has an initial contracting phase until the Universe reaches a minimum, followed by a bounce and the standard expanding phase after that. A cosmic bounce hence avoids the problem of initial singularity and associated problems present in inflation. A successful model of cosmic bounce, like in the case of inflation must solve the cosmological puzzles along with providing a mechanism for the generation of primordial density perturbations with a nearly scale-invariant spectrum. However, the most important feature of cosmic bounce is that the initial singularity present in the hot Big Bang universe is avoided. Although cosmological bounce is thus very appealing, a decreasing Hubble parameter around the bouncing point enabling the Universe to enter the phase of expansion from the contracting phase requires the violation of null energy condition. Hence to construct a model of bounce which suffers neither from ghosts nor gradient instabilities around the bouncing point is an extremely challenging task. A model of bounce which does not suffer from these problems and driven by the Standard Model~(SM) Higgs field would be of great interest. However, such a model could be realized only in modified gravity. Also, in Ref~\cite{biswas2012stable}, it was noted that Einstein gravity inflation cannot alleviate the Big Bang singularity problem and one must consider modified gravity. %\textcolor{blue}{cite[majumdar]}. 

Since early Universe physics takes place at high energies, one can't ignore the quantum corrections to gravity which should results in modifications to Einstein's general relativity~(GR). One of the possible extensions to Einstein's GR is the addition of Gauss-Bonnet~(GB) term to the Einstein Hilbert action. Gauss-Bonnet Gravity is a part of the general extension of gravity theories referred to as Lovelock theories of gravity~\cite{lovelock1971einstein,padmanabhan2013lanczos}. One key aspect of Lovelock theories is that the gravity equations of motion remain second order (and quasilinear in second order), unlike the~$f(R)$ gravity theories.  The Gauss-Bonnet term is the highest order Lovelock invariant in 4-D. However, since the Gauss-Bonnet term is a total derivative and non-dynamical in 4-D one must consider non-minimal coupling with the GB to make it dynamical. In Ref.~\cite{mathew2016low} the authors had considered a model of cosmic inflation driven by the SM Higgs field. In this paper, we show that the inflationary solution could have been preceded by a phase of contraction leading to a viable model of bounce inflation. In a model of bounce inflation, a process of bounce occurs before the traditional inflation with a graceful exit in the early Universe~\cite{wan2015bounce,bamba2016bounce,qiu2015g}.
 
The paper is organised as follows. In the next section, we present an exact solution for the phases of inflation and contraction. The exact solution for the inflationary evolution was first introduced in~\cite{mathew2016low}. Further in the section, we describe the evolution of the Hubble horizon in the two phases. In Sec.~\ref{sec:Numerical}, it is shown numerically that the phase of contraction can be joined to the phase of inflation through a bounce. Finally, in Sec.~\ref{sec:observations}, the observational parameters derived from our model are matched with observations. In the last section, we present  the discussions and conclusions.

We use~(-,+,+,+) metric signature and natural units,~$c=1,\quad \hslash=1$ and $\kappa\equiv{1}/{\MPlh^2}$, where $\MPlh$ is the reduced Planck mass. Lower Latin alphabets denote the 4-dimensional space-time, and lower Greek letters are used for the 3-dimensional space. Dot represents the derivative with respect to the cosmic time $t$, while prime denotes the derivative with respect to the conformal~time~$\eta$.  $H\equiv{\dot{a}}/{a}$ is the Hubble parameter.
\section{Exact solution}
\label{sec:exact_solution}
The action which is of interest is
{\small
\begin{eqnarray}  
{\cal S}=\int d^4x \sqrt{-g} \left[\frac{R}{2\kappa} +
f(\phi)\mbox{GB}-\frac{1}{2} \nabla_{a} \phi \nabla^{a} \phi -V\left(\phi\right)\right],%+{\cal S}_m
\label{eq:action}
\end{eqnarray} }
where~$\phi$ is a scalar field with the potential~$V(\phi)$, R~denotes the Ricci~scalar, and GB is the Gauss-Bonnet term given by
\begin{equation}
\mbox{GB}=R_{a b c d}R^{a b c d}-4R_{a b}R^{a b}+R^2
\end{equation}
In a spatially flat Friedmann-Robertson-Walker~(FRW) Universe 
\begin{equation}
ds^2=-dt^2 + a^2(t)(dx^2+dy^2+dz^2)
\end{equation}
with~$a(t)$ as the scale factor, the background equations are given by:
\begin{subequations}
\label{eq:threeequations}
\begin{eqnarray}
\label{eom}
 - 24 H^2 \frac{\ddot{a}}{a} {f}_{,\phi} \left(\phi\right) +
  \ddot{\phi}+V_{,\phi}\left(\phi\right)+ 3 H \dot{\phi} &=& 0 \\ 
%%%
\label{energyconstraint}
- 3 H^2 \left(\frac{1}{\kappa} +   8 H \dot{f}\left(\phi\right) \right) 
+ \frac{1}{2}\dot{\phi}^2 + V\left(\phi\right) &=& 0 \\
%%%
\label{evolutionequation}
-H^2 \left( \frac{1}{\kappa} + 8 \ddot{f}\left(\phi\right) \right)
 +V\left(\phi\right) \quad \quad \quad \quad \; & &\nonumber \\ 
\quad - \frac{2 \ddot{a}}{a} \left( \frac{1}{\kappa} +  8 H \dot{f}\left(\phi\right) \right)-\frac{1}{2} \dot{\phi}^2&=& 0
\end{eqnarray}
\end{subequations}
where H is the Hubbles parameter $\dot{a}/a$.

In Ref.~\cite{mathew2016low}, it was shown that for the above action there exists an exact solution
\begin{subequations}
\label{eq:soln}
\begin{eqnarray}
a(t) &=& a_0 (t +C_1)^p\\
\phi(t) &=& \phi_0 (t + C_1)^{-\frac{1}{2}}
\end{eqnarray}
\end{subequations}
 for potential and coupling function
\begin{subequations}
\label{eq:PLsplsol}
\begin{eqnarray}
V\left(\phi\right)&=& \lambda_4  {\phi}^{4}  + \lambda_6 \phi^6 \\
 %%%%%%%%%%%%%%%%%%%%     
 f\left(\phi\right)&=& \frac{\alpha_2}{\phi^2} +  \frac{\alpha_4}{\phi^4} 
\end{eqnarray}
\end{subequations}
where $p>1$ for a positive quartic coupling~($\lambda_4>0$) and the coefficients are given by 
\begin{eqnarray}
%\begin{multilined
\label{eq:PLcoefficients}
\lambda_4  &=& {\frac {3{p}^{2} \left(p-1 \right) }{{{ \phi_0}}^{4}\kappa
    \left( p+1 \right) }} \quad
\lambda_6 = \frac{1}{4}\,{\frac { \left( 5\,p-2\right)}{{{ \phi_0}}^{4} \left( 4+2\,p
    \right) }} \quad\nonumber \\ \nonumber \\ \nonumber\\
\alpha_2 &=&      \frac{1}{32}\,{\frac{{{\phi_0}}^{4}}{{p}^{2}
    \left( 2+p \right) }}  \quad
\alpha_4 = - \frac{1}{8}\,{\frac {{{\phi_0}}^{4}}{\kappa p \left( p+1 \right) }}  
%\end{multilined}     
\end{eqnarray}
From above relations one can see that the constant~$\phi_0$ can be written in terms of the coefficients~$\lambda_4$~and~$\alpha_4$ as
\begin{equation}
\label{eq:phi0incoeffs}
\phi_0^2  =  \left( 24 p^3 (1-p) \frac{\alpha_4}{\lambda_4}  \right)^{1/4} 
\end{equation}
Also, $\alpha_4$ can be written in terms of $\lambda_4$ and $p$ as:
\begin{equation}
\label{eq:alphainlambdap}
\alpha_4=-\frac{3}{8\lambda_4} \frac{p(p-1)}{\kappa^2(p+1)^2}
\end{equation}
The power-law solution given in Eq.~\ref{eq:soln} is valid only for $t>0$ as the scalar field becomes imaginary for $t<0$. 
The field equations Eq.~\ref{eq:threeequations} are invariant under time reversal and time translation. Hence, if Eq.~\ref{eq:soln} is one solution of the model then $t\rightarrow -t+C$ where $C$ is any arbitrary constant will also yield a solution which is valid only for $t<0$ again for the same reason that the scalar field will go imaginary but here for $t>0$.
% %

From the above discussions it is clear that another solution for the field equations given below also satisfies the field equations for the potential and coupling function given by Eq.~\ref{eq:PLsplsol}. Also, we have explicitly verified that indeed Eq.~\ref{eq:gensoln} is a solution of the model.
%The power-law solution given in Eq.~\ref{eq:soln} is valid only for $t>0$ as the scalar field becomes imaginary for $t<0$. However, we know that if Eq.~\ref{eq:soln} is one solution of the model then $t\rightarrow -t+C$ where $C$ is any arbitrary constant will also yield a solution which is valid only for $t<0$ again for the same reason that the scalar field will go imaginary but here for $t>0$. 

%From the above discussions it is clear that Eq.~\ref{eq:gensoln} given below also satisfies the field equations for the potential and coupling function given by Eq.~\ref{eq:PLsplsol}.
\begin{subequations}
\label{eq:gensoln}
\begin{eqnarray}
a(t) &=& a_0 (\pm t +C_2)^p\\
\phi(t) &=& \phi_0 (\pm t + C_2)^{-\frac{1}{2}}
\end{eqnarray}
\end{subequations}
Now as explained in Ref.~\cite{mathew2016low}, for  $\phi\ll \MPlh$, the potential and coupling of form
\begin{equation}
\label{Eq:Higgslikepot}
V(\phi)=\lambda_4\phi^4 \quad \mbox{and} \quad f(\phi)=\alpha_4 \phi^{-4}
\end{equation}
lead to the evolution of the universe approximately given by Eq.~\ref{eq:soln}. $\lambda_4$ is an arbitrary constant and can take any value we want. This motivates us to consider the scalar field to be the Higgs field i.e., we assume $\lambda_4=\lambda/4$, where $\lambda$ is the Higgs quartic coupling. From here onwards, we consider the action
\newline
{\small
\begin{equation}
\label{eq:actionHiggs}
S = \int d^4 x \sqrt{-g} \left[\frac{R}{2\kappa} + \frac{\alpha_4}{h^4} {GB} - \frac{1}{2} g^{a b} \partial_a h \partial_b h 
- \lambda_4 \left(h^2 - v^2\right)^2 \right], 
\end{equation}}
\newline
where $h$ is the Higgs field and $v$ is the vacuum expectation value of the Higgs field.

One can easily see that the solution  
\begin{equation}
\label{eq:contraction}
a(t)=a_0(-t+C_1)^p \quad \mbox{and} \quad h(t)=h_0(-t+C_1)^{-\frac{1}{2}}
\end{equation}
leads to a contracting Universe. Also, for such an evolution~(for $p>1$) the equation of state~($\omega)$ is less than one, i.e., the solution does not lead to the most appealing ekpyrotic bounce~\cite{khoury2001ekpyrotic,buchbinder2007new}. However, it must be noted that in the case of single scalar field models it is extremely difficult to construct ekpyrotic models of bounce that give scale-invariant power-spectrum. An extensively studied model of bounce which is not an ekpyrotic attractor is that of matter bounce, where the Universe contracts at early times as in a matter dominated epoch. The matter bounce is known to generate a scale-invariant spectrum. 
\begin{figure} \centering
\includegraphics[width=15pc]{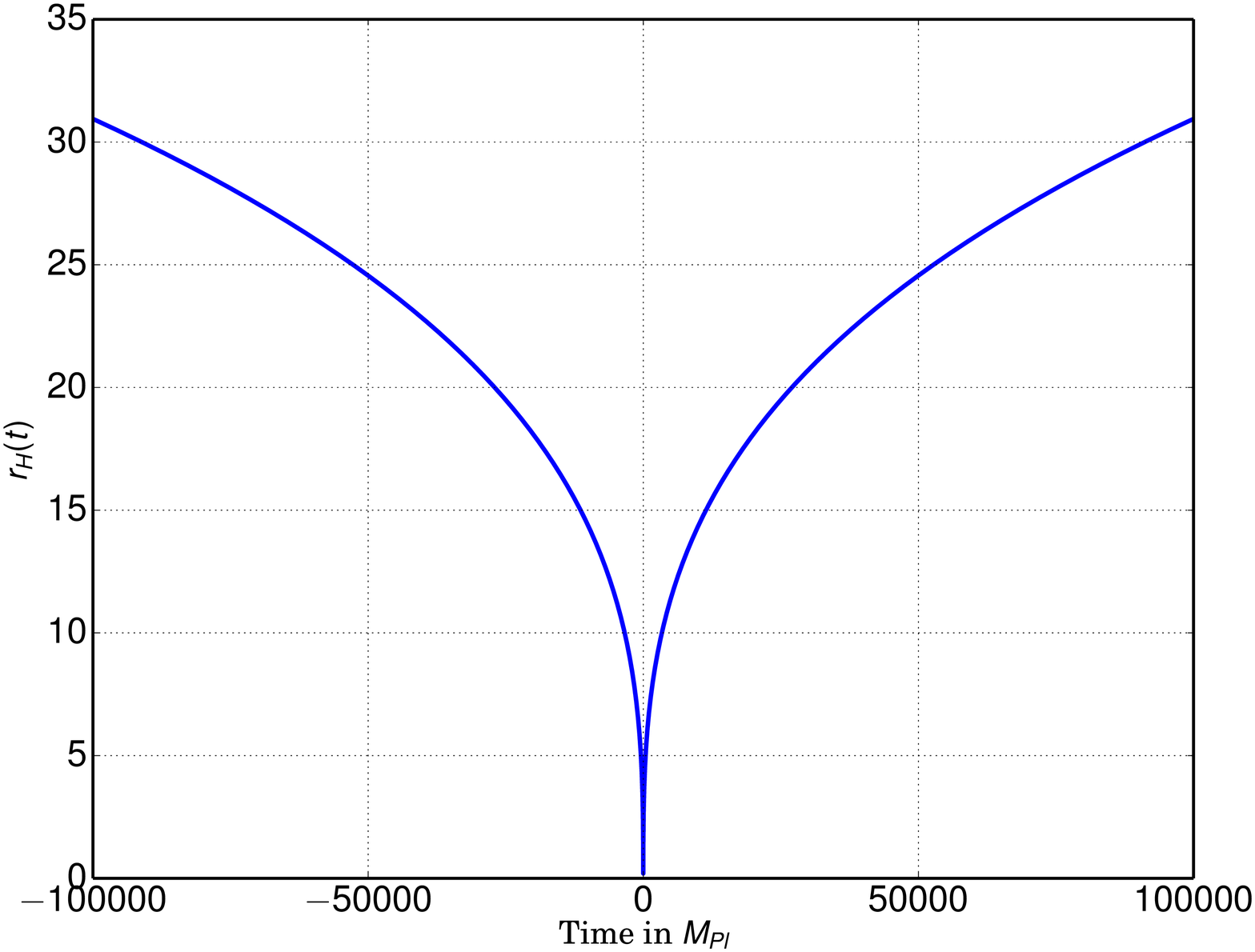}
\includegraphics[width=15pc]{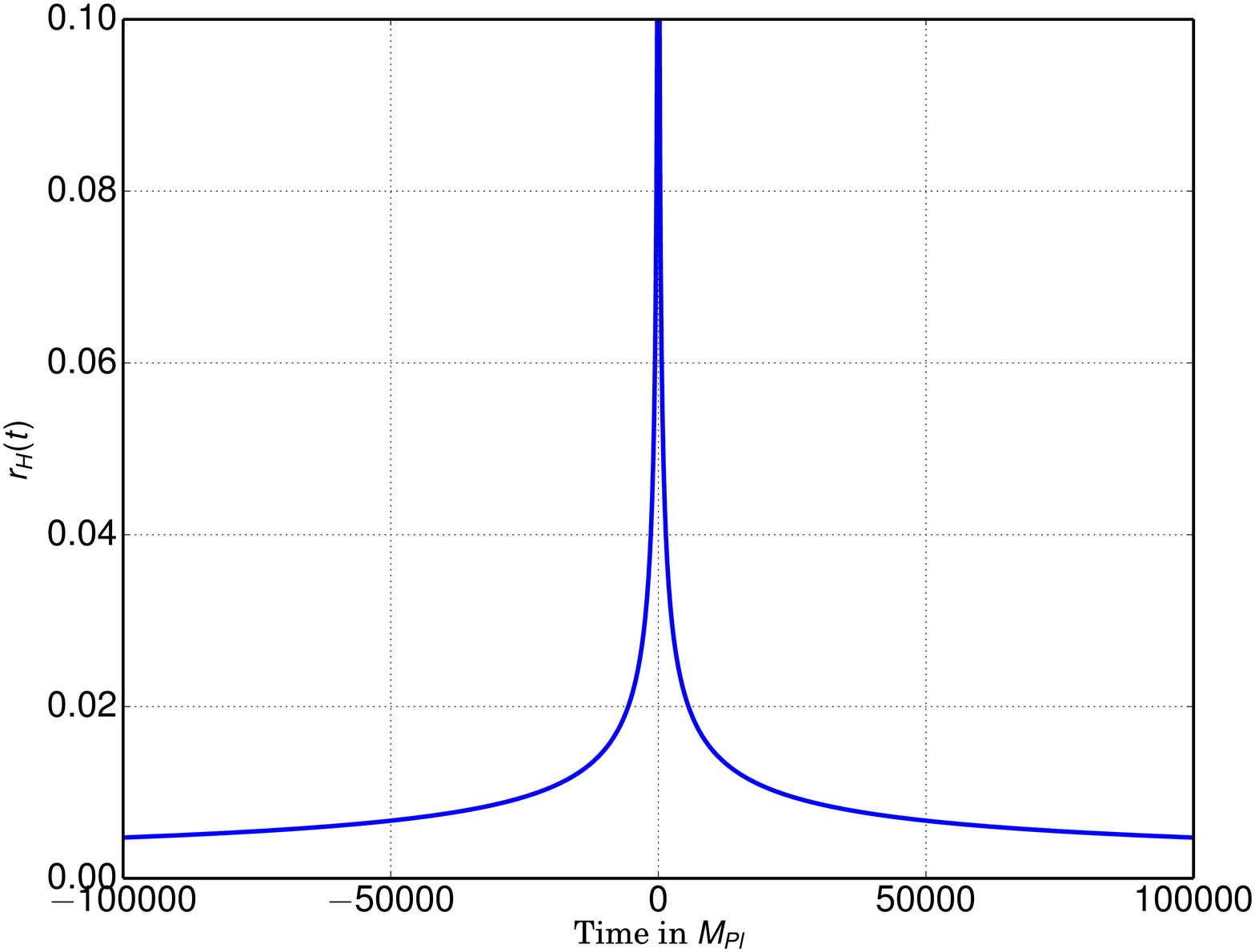}
\caption{The comoving Hubble radius $r_H(t)$ for the matter bounce
scenario (above)%(left plot)
 and for the power-law contraction scenario (below). % (right plot).
 } \label{fig:horizoncomp}
\end{figure}
%\twocolumngrid
 %In the case of bouncing scenarios where the comoving horizon decreases during the contraction the Bunch-Davies initial condition for the perturbed modes are assumed at extreme past, i.e., when the modes are inside the horizon. Unlike this and like in Ref.~\cite{odintsov2017big}, in our model the Bunch-Davies initial conditions are assumed around or after the bouncing point.
 
 In the case of models of bounce considered heavily in literature including that of matter bounce, the Hubble horizon decreases during contraction. Hence, the observationally relevant cosmological modes originate at extreme past compared to the bouncing point and the cosmological perturbations are generated during the phase of contraction. During the era of contraction, for the case of matter bounce these primordial modes were at sub-horizon scales since the Hubble horizon shrinks during contraction from an infinite size. Also, for these models the Bunch-Davies initial condition for the perturbed modes are assumed at extreme past, i.e., when the modes are inside the horizon. 
 
 In sharp contrast, the physical picture is totally different in our model. In fact, in our model the Hubble horizon increases during the contracting phase and the Hubble horizon has an infinite size only for cosmic times about the bounce (see Fig.~\ref{fig:horizoncomp}, where the evolution of Hubble horizon for a matter bounce and our model, power-law contraction,  $p>1$, is given). Hence the primordial modes that are relevant for present time era are generated near the bouncing point, because it is about the period close to bounce that the primordial modes are contained
 within the Hubble horizon. As the horizon shrinks during
 the epoch of inflation following the bounce, these modes
 exit the horizon and are relevant for the
 present time observations. In Ref.~\cite{odintsov2017big}, the authors consider such a scenario where the comoving horizon increases during the
 phase of contraction. The authors assumes Bunch-Davies
 initial conditions for perturbations around the bouncing
 point assuming the Fourier modes relevant for present time observations to have been generated near the bouncing point. The physical picture of our model is similar to Ref.~\cite{odintsov2017big} and the Bunch-Davies initial conditions for the perturbed modes are assumed around the bouncing point.  
 
 The expanding part of our model is identical to the standard inflationary scenario. The difference from the standard scenario of inflation following the big bang being the fact that there exists a phase of contraction and a bounce and the scale factor never goes to zero. Also compared to the standard bouncing scenarios, the relevant modes observed in the CMB are generated around the bouncing point and leave the horizon during the inflationary epoch.  In the next section, we present our numerical results showing that the contracting phase is followed by a bounce after which a normal inflationary evolution follows with exit.
%%%%%%%%%%%%%%%%%%%%%%%%%%%%%%%%  
\section{Numerical Results:}
\label{sec:Numerical}
We know that the contracting phase is not an attractor since the equation of state ($\omega$) is less than one. Also, we know from Ref.~\cite{mathew2016low} that the expanding/inflationary phase is an attractor. This means that the solution obtained by inverting time (contracting solution) has to be a repeller (An attractor solution is a repeller backwards). This means that the initial value of $\dot{h}$ has to be close to the corresponding value of $\dot{h}_{ES}(t_i)$ for the power-law exact solution, the initial value of Hubble parameter is constrained by the energy constraint, so that the Universe stays in the contracting phase. This is one drawback of our model, however, most bouncing solutions including matter bounce suffers from this problem. The early Universe models including inflation and bounce is important not just because it solves the cosmological puzzles but for its ability to provide a successful mechanism for the generation of initial density perturbations leading to the formation of structure. Now we have seen that the contracting solution is a repeller and the expanding or inflationary solution to be an attractor. In this section we show numerically that for a wide range of initial conditions with deviations in $\dot{h}(t_i)$, the Universe stays in the contracting phase for a while and then smoothly transit into the attractor solution (inflationary solution). 

We have solved the equations of motion~\ref{eq:threeequations} numerically for $\lambda_4=0.032275$ (using the LHC results for Higgs field, quartic coupling for Higgs field $\lambda=0.129$) and $\alpha_4=-11.054$. These values for the coupling constants corresponds to $p=60$ given from Eq.~\ref{eq:alphainlambdap} and $h_0=23.9$ given from Eq.~\ref{eq:phi0incoeffs}. The constant $C_1$ appears through the initial value of the field ($h(t_i)$) and for our calculations we have used $h(t_i)=h_0 (t_i)^{-1/2}$ assuming $C_1=0$. The initial velocity of the field, $\dot{h}(t_i)$ is allowed to slightly deviate from that of the exact solution ($\dot{h}_{ES}(t_i)$).

\begin{figure}[!htb] \centering
\includegraphics[width=0.45\textwidth ]{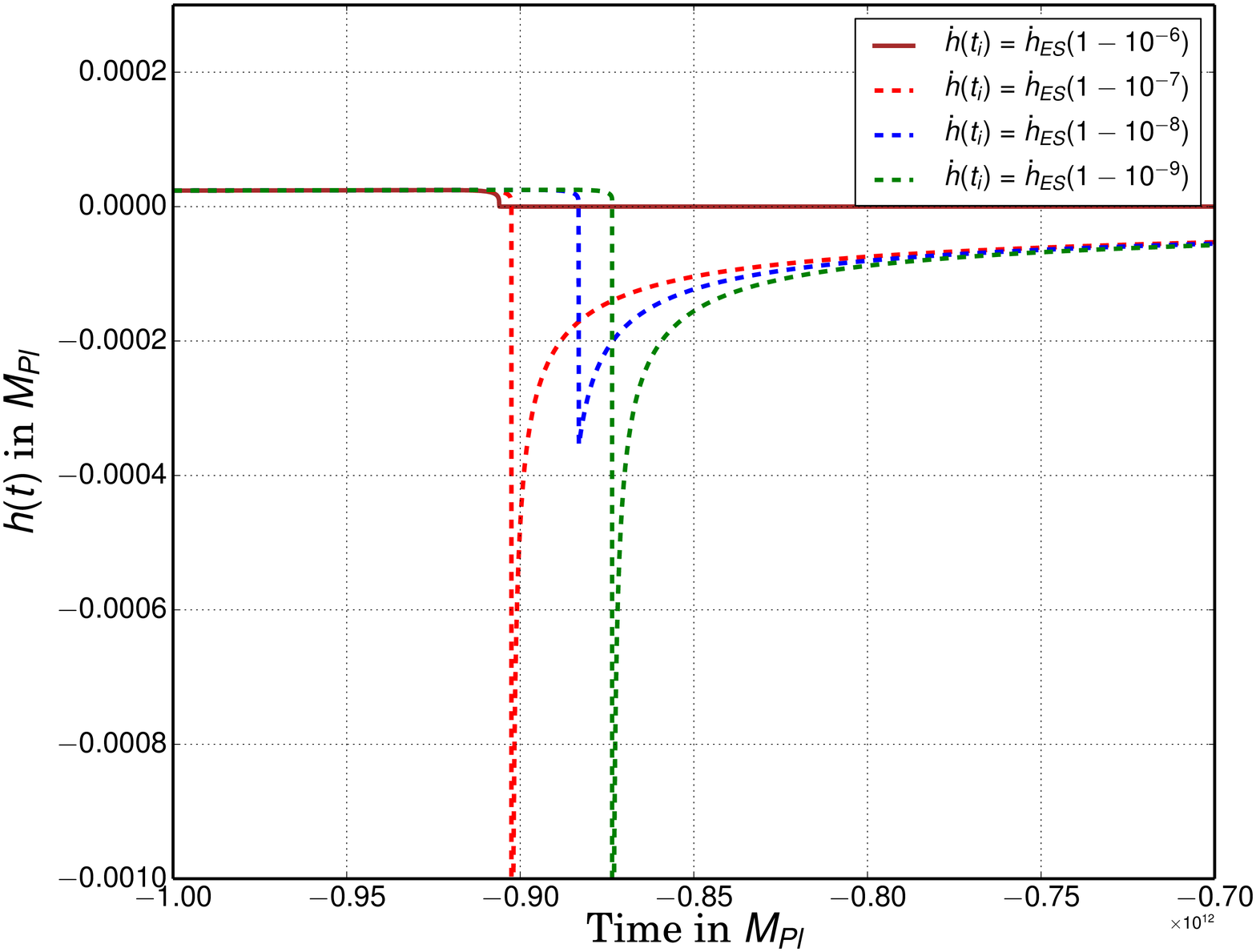}
\caption{The evolution of the Higgs field for different initial conditions depicting the transition from the contracting phase to the expanding phase.
 } \label{fig:scalarfield}
\end{figure}
\begin{figure}[!htb] \centering
\includegraphics[width=0.45\textwidth ]{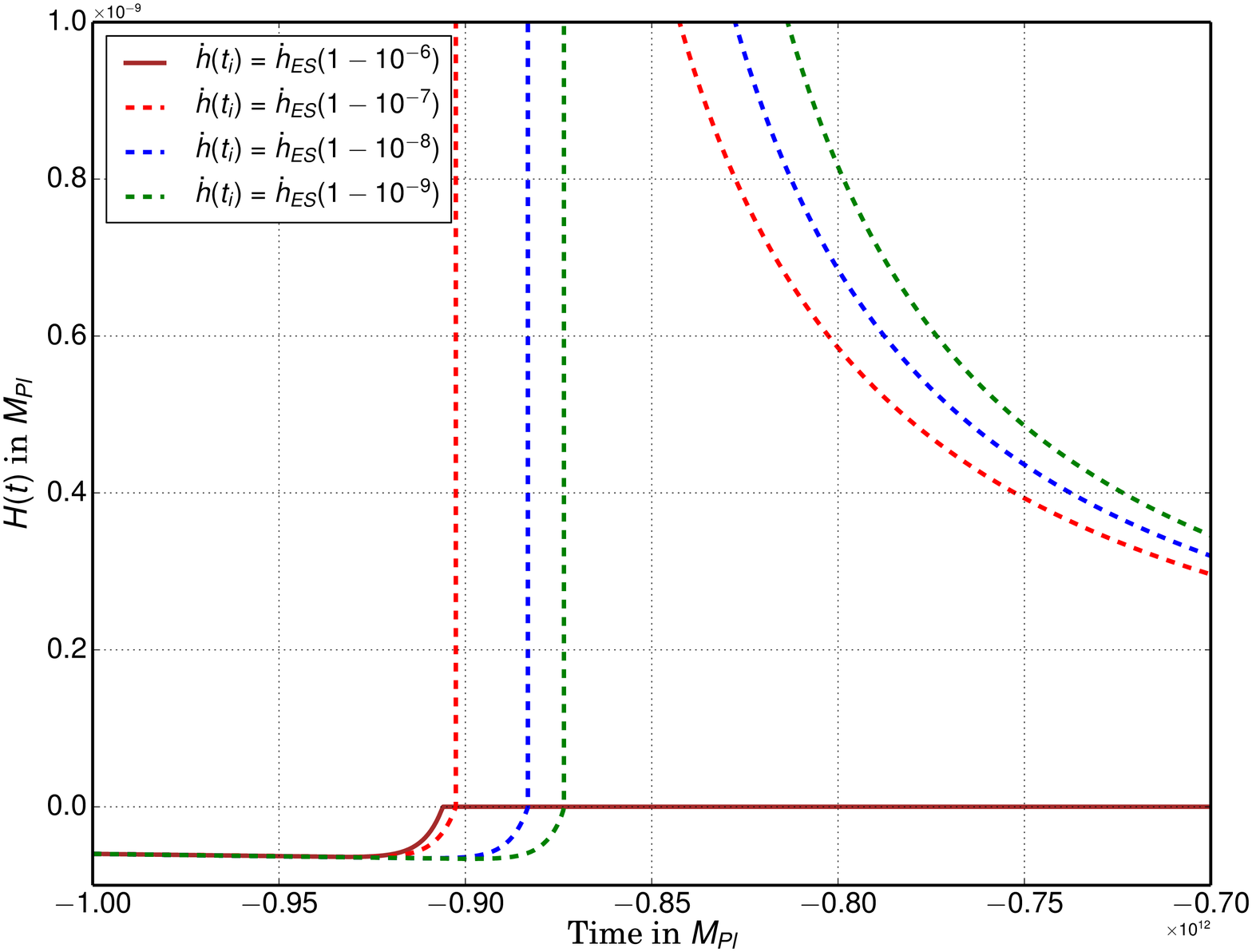}
\caption{The evolution of the Hubble parameter for different initial conditions depicting the transition from contracting phase to expanding phase.
 } \label{fig:hubbleparameter}
\end{figure}
\begin{figure}[!htb] \centering
\includegraphics[width=0.45\textwidth ]{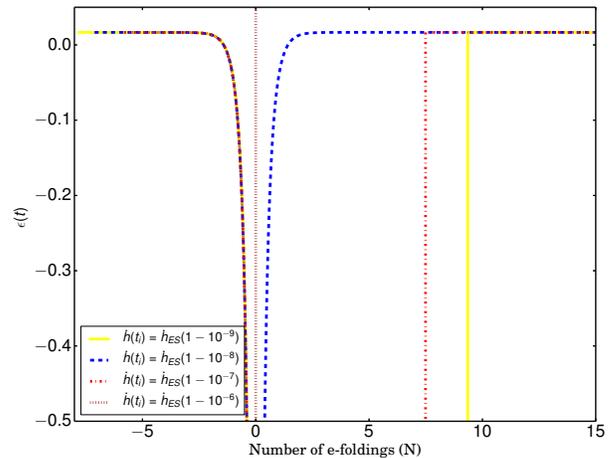}
\caption{The evolution of the slow-roll parameter for different initial conditions depicting the transition from contracting phase to expanding phase.
 } \label{fig:slowroll}
\end{figure}

In Fig.~\ref{fig:scalarfield}, we have plotted the evolution of the Higgs field $h$ with respect to time. It is clear from the graphs that the repeller contracting solution smoothly transit to the inflationary solution. The brown curve in the plot is where the Universe instead of entering into the expanding phase enters into a static phase.  In fig~\ref{fig:hubbleparameter}, we have plotted the Hubble parameter (H) with time. It must be noted that for our model the maximum value of both Hubble parameter and the Higgs potential are well below Planck energy. 

A quantity that captures the essence of the shift of solution is slow-roll parameter:
\begin{equation}
\epsilon = -\frac{\dot{H}}{H^2}
\end{equation}
because $\epsilon=1/p$ and is independent of the constant $C_1$ (for power-law solution) for both the era of contraction and the era of inflation. We have plotted the slow-roll parameter vs number of e-foldings in fig~\ref{fig:slowroll}. See that $\epsilon$ takes the value $\epsilon = 1/60$ for $p=60$ during the initial contracting phase and in the final inflationary epoch where the Universe remains in the exact solution. The smooth transition from the era of contraction to that of inflation is captured in the plot of the evolution of the Higgs field. One important thing to note is that the value of the constant $h_0$ gets its sign inverted (Note that from Eq.~\ref{eq:phi0incoeffs}, we know that $h_0$ can take both positive and negative values for the same coefficients) and the entire solution is time translated, i.e., $C_1$ is no longer zero but takes an arbitrary value depending on the deviation in initial scalar field velocity. From the fact that both $p$ and $h_0$ are fixed by the values of the coupling constants and from the plots of slow-roll parameter and Higgs field, one can see that the inflationary solution is:
\begin{subequations}
\begin{eqnarray}
h(t)&=& -h_0 (t+C_2)^{-1/2}\nonumber \\
a(t)&=& a_0 (t+C_2)^p\nonumber
\end{eqnarray}
\end{subequations}
where $C_2$ is arbitrary and depends on the initial velocity of Higgs field. 

\section{Matching with Observtions}
\label{sec:observations}
As we have mentioned earlier, in our model of bounce inflation the perturbed modes leave the horizon during the phase of expansion and the calculations remain similar to that in the case of inflation. Hence,the  calculations of the power-spectrum for our model are given in Ref.~\cite{mathew2016low}. Here we have given only the relevant results.

The scalar power-spectrum is given by: 
\begin{eqnarray}
\label{eq:ScalarPowerSpectrum}
P_{\cal R}&=& {\cal A}_s \, k^{3 - 2\sigma};  \\
{\cal A}_s &=& 2^{4(\sigma - 1)}c^{-2\sigma} \left(\frac{\Gamma(\sigma)}{\Gamma(3/2)}\right)^2\frac{1}{4 \pi^2} \left(\frac{a_0}{2\sigma-3}\right)^{2\sigma -1} \frac{1}{a_0^2 Q_{\mathcal{R}}}  \nonumber
\end{eqnarray}
where 
\begin{equation*}
 \sigma=\frac{3p-1}{2(p-1)} \quad and \quad Q_\mathcal{R}=\approx~~0.0017~\MPlh^2.
 \label{Qs}
 \end{equation*}
 $c$ is obtained to be a constant with a value slightly less than $1$. The tensor power-spectrum for the model is:
 \begin{eqnarray}
 \label{eq:TensorPowerSpectrum}
 P_{\cal{T}}&=& {\cal A}_T k^{3-2\sigma} \\ 
 {\cal A}_T &=& \frac{8\times 2^{4(\sigma -1)}c_{\cal{T}}^{-2\sigma} }{a_0^2 Q_{g}}\left(\frac{\Gamma(\sigma)}{\Gamma(3/2)}\right)^2\frac{1}{4\pi^2} \left(\frac{a_0}{2 \sigma-3}\right)^{2 \sigma -1} 
 \nonumber
 \end{eqnarray}
 where 
 \begin{equation}
  Q_g=\frac{1}{\kappa}+8H\dot{f},\quad c_{\cal{T}}^2 = \frac{1 +8 \kappa \ddot{f}}{1 + 8 \kappa H \, \dot{f}} \, .
 \end{equation}
 We can see that both $Q_g$ and $c_{\cal{T}}$ are both constants, where $c_{\cal{T}}\approx 1$.
 The scalar spectral index $n_{\cal{R}} -  1= 3 - 2 \sigma$. From Planck constraint~\cite{akrami2018planck}, we know that the scalar spectral index is given by approximately $0.965\pm0.004$. This implies that the constant $p$ takes values around 60. Since, we have assumed the scalar field to be Higgs field the coupling coefficient $\lambda_4=\lambda/4$ where $\lambda=0.129$, the observed value of the Higgs quartic coupling. This fix the parameter $\alpha_4$. All the parameters of the model are fixed and  the tensor to scalar ratio is obtained to be $0.012$.
 \section{Discussions}
 Bounce inflation is an interesting paradigm for early
 Universe where the singularity problem associated with
 inflationary paradigm doesn't appear. Furthermore, a model
 of bounce inflation retains all the advantages associated
 with inflation. In this paper, we present a model of bounce
 inflation driven by the Higgs field where the Higgs field is non-minimally coupled with gravity through a Gauss-Bonnet term. Our paper is an extension of Ref.~\cite{mathew2016low} in which
 a model of inflation driven by the Higgs field was considered.
 Like in Ref.~\cite{mathew2016low}, in this model the exit from inflation happens at electro-weak scale~\cite{knox1993inflation,garcia1999nonequilibrium,krauss1999baryogenesis,german2001low,van2004electroweak,ross2016hybrid}, and a non-zero Higgs mass is crucial in obtaining the graceful exit. However, we haven't discussed regarding the exit in detail as once the Universe falls into the inflationary regime everything remains the same as in Ref.~\cite{mathew2016low}. Our model is consistent with the observations predicting a viable scalar spectral index for a suitable value of the coupling co-efficient and a small value for tensor to scalar ratio as required
 by the CMB observations. Our model requires fine-tuned initial conditions. However, the fine-tuning associated
 with the model is not as bad as the fine-tuning associated with the hot big bang model. Furthermore, the most appealing feature of the early Universe models such as inflation and bounce are not solving the cosmological puzzles but providing a successful mechanism for the generation of initial density perturbations seeding the formation of structure. A successful model of bounce requires the violation of the null energy condition. This makes building models of bounce without ghost or gradient instabilities an extremely challenging task. In our work we have given a viable model of bounce inflation driven by Higgs field. 
\section{Acknowledgements}

Jose~Mathew wish to thank Sreedhar~B~Dutta, S.~Shankarnarayanan and Krishnamohan~Parattu for useful discussions and suggestions.
%JM thank Dr. Krishnamohan Parattu  for valuable suggestions and discussions. JM thank Prof. L. Sriramkumar for supporting him during his visit at IITM.
%\clearpage
%%%%%%%%%%%%%%%%%%%%%%%%%%%%%%%%
%\newline
%\linebreak
%\begin{center}{\noindent\rule{15cm}{0.4pt}}\end{center}
%\bibliographystyle{unsrtnat}
%\bibliography{bibdata.bib,bibdata1.bib,bibdata2.bib,bibdata3.bib}

\end{document}